\documentclass[preprint,12pt, a4paper]{elsarticle}

\usepackage{amssymb}
\usepackage{amsmath}
\usepackage{lineno}
\usepackage{color}
\usepackage{soul}
\usepackage{comment}
\usepackage{caption}
\usepackage{float}
\usepackage{wasysym}
\restylefloat{table}
\usepackage{booktabs}
\usepackage{tabularx}
\usepackage{textcomp}
\usepackage{placeins}
\usepackage{hyperref}
\usepackage[margin=3cm]{geometry}
\usepackage{setspace}
\hypersetup{
	colorlinks = true,
	urlcolor = blue,
	linkcolor = blue,
	citecolor = red
}

\biboptions{sort&compress}

\journal{Additive Manufacturing}

\begin{document}
	
\begin{frontmatter}



\title{\textbf{Effect of hirtisation on the roughness and fatigue performance of porous titanium lattice structures}}


\author[label1,label2]{Reece N. Oosterbeek \corref{cor1}}
\ead{reece.oosterbeek@eng.ox.ac.uk}

\author[label3]{Gabriela Sirbu}

\author[label3]{Selma Hansal}

\author[label4]{Kenneth Nai}

\author[label1]{Jonathan R. T. Jeffers}

\cortext[cor1]{Corresponding author}

\address[label1]{Department of Mechanical Engineering, Imperial College London, London, SW7 2AZ, United Kingdom}
\address[label2]{Department of Engineering Science, University of Oxford, Parks Road, Oxford OX1 3PJ, United Kingdom}
\address[label3]{RENA Technologies Austria GmbH, Samuel-Morse-Strasse 1, 2700 Wiener Neustadt, Austria}
\address[label4]{Renishaw PLC, New Mills, Wotton-under-Edge, Gloucestershire GL12 8JR, United Kingdom}

\begin{abstract}

Additive manufacturing (AM) has enabled the fabrication of extremely complex components such as porous metallic lattices, which have applications in aerospace, automotive, and in particular biomedical devices. The fatigue resistance of these materials is currently an important limitation however, due to manufacturing defects such as semi-fused particles and weld lines. Here Hirtisation\textsuperscript{\circledR} is used for post-processing of Ti-6Al-4V lattices, reducing the strut surface roughness (\textit{Sa}) from 12 to 6 \textmu m, removing all visible semi-fused particles. The evenness of this treatment in lattices with $\rho /\rho_{s}$ up to 18.3\% and treatment depth of 6.5 mm was assessed, finding no evidence of reduced effectiveness on internal surfaces. After normalising to quasi-static mechanical properties to account for material losses during hirtisation (34-37\% reduction in strut diameter), the fatigue properties show a marked improvement due to the reduction in surface roughness. Normalised high cycle fatigue strength ($\sigma_{f,10^{6}}/\sigma_{y}$) increased from around 0.1 to 0.16-0.21 after hirtisation, an average increase of 80\%. For orthopaedic implant devices where matching the stiffness of surrounding bone is crucial, the $\sigma_{f}/E$ ratio is a key metric. After hirtisation the $\sigma_{f}/E$ ratio increased by 90\%, enabling design of stiffness matched implant materials with greater fatigue strength. This work demonstrates that hirtisation is an effective method for improving the surface roughness of porous lattice materials, thereby enhancing their fatigue performance.

\end{abstract}


\end{frontmatter}

\FloatBarrier
\begin{figure}[h!]
	\centering
	\includegraphics[width=130mm]{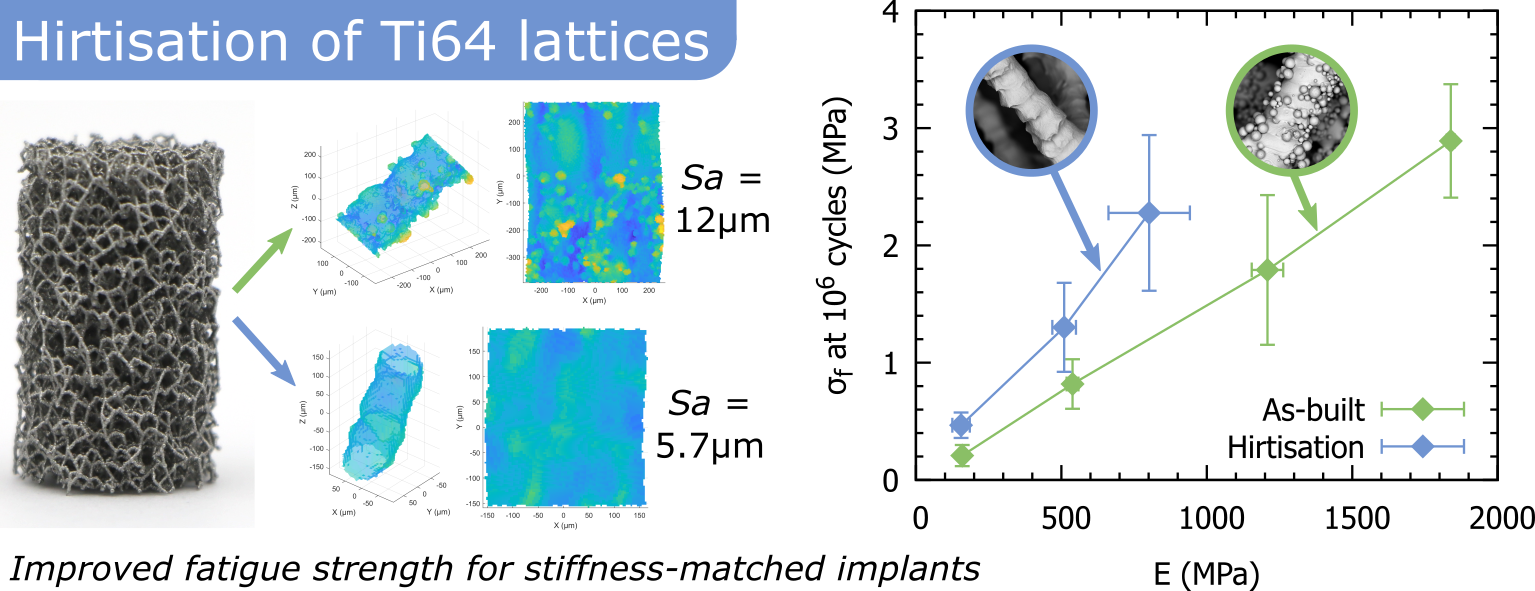}
	\caption*{Graphical abstract}
\end{figure}
\hrulefill
\FloatBarrier



\section{Introduction}
\label{}

Porous lattice materials fabricated by metal additive manufacturing (AM) are now realistic engineering choices for a large variety of applications including lightweight aerospace or automotive components \cite{Shapiro2016, Plocher2019} and energy absorbing structures \cite{Zhang2021}. For biomedical applications in particular these materials demonstrate considerable potential, with the design of pore networks enabling bone ingrowth into the porous lattice material \cite{Wang2016a, Zhang2019a, Ghouse2019}. In these applications the ability to controllably tune the material elastic modulus is a key advantage, allowing the material stiffness to be matched to the mechanical properties of the surrounding bone, minimising stress shielding and promoting bone ingrowth and implant fixation \cite{Ghouse2019, Sumner2015}. The fatigue behaviour of these porous lattice materials is a critical concern for any applications in biomedical implants, where normal service conditions involve highly cyclic loading for extended periods, with severe consequences for material failure. This is a particular weakness for lattice materials fabricated by additive manufacturing, which are characterised by high surface area, a large quantity of surface defects, and non-equilibrium microstructures \cite{Thijs2010, Gong2014, Charles2019}. To balance fatigue resistance with stiffness-matching demands, maximising the $\sigma_{f}/E$ ratio is a critical design challenge for engineering of porous lattices for biomedical implants.

Metal lattices produced by powder bed fusion (PBF, a prevalent metal AM technique) commonly display surface defects such as weld lines between layers, and semi-fused particles, increasing the surface roughness and likelihood of fatigue crack initiation \cite{Bagehorn2017}. The rapid cooling rates experienced during PBF also lead to the presence of significant residual stresses, and formation of non-equilibrium microstructures such as a martensitic $\alpha'$ phase in Ti-6Al-4V, resulting in poor resistance to fatigue crack propagation \cite{Xu2015, Li2016b}. This can be alleviated by heat treatment procedures such as hot isostatic pressing or vacuum heat treatment, leading to improvements in the fatigue lifetime \cite{Ghouse2021, Wauthle2015a}. Treatments to improve the lattice surface roughness are often more challenging, due to the difficulty of evenly treating both internal and external struts \cite{Wysocki2019}. Chemical etching techniques, often using a combination of HF and HNO\textsubscript{3} have been studied, demonstrating reduced roughness and improved fatigue performance \cite{DeFormanoir2016, Lhuissier2016, Persenot2020, Pyka2013}. However, many such studies focus on representative single struts, and the applicability of these results to larger lattices may be limited, due to the restriction of fluid flow through the bulk lattice as demonstrated by Wysocki et al. \cite{Wysocki2019}. Further work by Karami et al. used a combination of light etching and sand blasting of strut-based lattices, demonstrating qualitative improvements in surface roughness resulting in improved fatigue strength \cite{Karami2020}.

To overcome some of the limitations with conventional chemical etching, methods such as Hirtisation\textsuperscript{\circledR}\footnote{Hirtisation\textsuperscript{\circledR} is a registered trademark of RENA Technologies GmbH, and will hereafter be referred to as hirtisation.} have been developed, incorporating dynamic electrochemical and hydrodynamically supported chemical processes to combine the benefits of both process routes during the different steps of post-processing \cite{Hansal2023}. Recent work by Sandell et al. investigated the surface roughness of AM Ti-6Al-4V monolithic parts treated by hirtisation, with successful roughness reduction by 75\% observed \cite{Sandell2022}. The combination of electrolyte composition (aqueous acidic or alkaline electrolytes), hydrodynamic flow, and dynamic electrochemistry (interrupted or pulsed current processes) allow tailoring of the effective current field lines and potential fields to specifically target surface roughness reduction. This process is hypothesised to provide even surface treatment throughout a porous lattice \cite{Hansal2023} and improve the fatigue performance, however to date this has not been reported on in the literature.

One barrier to assessing the effectiveness of different surface treatments for additively manufactured lattices is the difficulty of accurately quantifying changes in the surface roughness of internal struts. Established roughness measurement methods depend on line-of-sight and utilise optical or contact (stylus) profilometry to characterise near-flat surfaces, however these techniques are not feasible for internal features. As a result, 3D X-ray imaging using micro-CT is typically used. Despite this, extracting strut surface roughness data can still be challenging, with methods involving extracting line profiles \cite{Kerckhofs2012, Kerckhofs2013} and CAD variance \cite{DuPlessis2018} described in the literature. The software package StrutSurf, recently developed by the authors \cite{Oosterbeek2022}, aims to sample a lattice and fit 3D geometries to individual struts from micro-CT data. This provides detailed information on strut roughness and morphology across the lattice, allowing more comprehensive investigation of the effects of various surface treatments.

This work aims to quantitatively assess the effectiveness of hirtisation for reducing the surface roughness of AM Ti-6Al-4V porous lattices using micro-CT and new analysis methods, and determine the effect of these surface changes on the mechanical properties, in particular the high cycle fatigue response.

\section{Materials and methods}

\subsection{Lattice design}
Commercial CAD software (Rhinoceros 6.0, Robert McNeel \& Associates) was used to design the stochastic lattice structures, using a method described in detail elsewhere \cite{Ghouse2017, Hossain2021} which will be briefly summarised here. A cylindrical geometry was selected (\diameter 13 mm $\times$ 22 mm) with height/diameter ratio of $>$1.5 in accordance with ISO 13314 \cite{ISO13314}. A poisson disk algorithm was used to fill this geometry with randomly distributed points (nodes), which were then connected to achieve a desired connectivity ($Z = 5.8$) and strut density ($d = 4.1$ mm\textsuperscript{-3}). To avoid low angle struts which are difficult to fabricate, struts with a build angle of $<$25\textdegree\ to the build plate were kinked upwards (split in half and the midpoint shifted upwards) \cite{Hossain2021}. Build files (slice data) were generated using in-house software and known calibrated laser parameters for desired strut thicknesses \cite{Ghouse2017}.

\subsection{Materials and manufacturing}
A RenAM500Q metal powder bed fusion (PBF) system (Renishaw plc., UK) was used to produce the lattice samples, using a Grade 23 Ti-6Al-4V ELI alloy powder with spherical particles (size range 15-45 \textmu m). Laser scanning was carried out using a contour strategy, with contour diameter of 70 \textmu m, laser power of 50-166 W, point distance of 50 \textmu m, exposure time of 100 \textmu s, and layer thickness of 50 \textmu m. After fabrication samples were subjected to vacuum heat treatment to relieve residual stresses and remove the martensitic microstructure \cite{Ghouse2021}. Samples were heated in a VF1218H vacuum furnace (Specnow Ltd.) at 13\textcelsius/min under vacuum (10\textsuperscript{-5} mbar) to 850\textcelsius, held for 127 minutes, before furnace cooling to room temperature under vacuum. Samples were removed from the titanium build plate using electrical discharge machining, followed by grinding of the top and bottom surface to ensure parallelism. Before further testing samples were immersed in ethanol and cleaned in an ultrasonic cleaner. Throughout this article ``as-built" refers to samples that have been heat treated, removed from the build plate, ground, and cleaned as above, before surface treatment (hirtisation).

\subsection{Hirtisation}
Lattices with 4 different strut thicknesses were fabricated, and 3 of these were then subjected to further processing using hirtisation before sample removal from the build plate (not carried out on the lowest relative density lattice, $\rho /\rho_{s} = 7.3\%$, due to expected material losses). Hirtisation was performed by Rena Technologies Austria using a H3000 automated industrial finishing machine, optimised for post-processing for small-scale production. The electrolyte system Hirtisation Ti-Auxilex was used, which is specifically designed for Ti alloys such as Ti-6Al-4V. A one-step Titanium treatment was carried out at 25\textdegree C, with 100\% power level for 4 minutes. After treatment  samples were thoroughly rinsed with distilled water and vacuum dried at 50\textdegree C for 20 minutes.

\subsection{Material characterisation}
Micro-CT scans were carried out using a Zeiss Xradia 510 Versa X-ray microscope. The X-ray voltage and current used were between 80-140 kV and 72-88 \textmu A respectively, depending on sample transmission. 2401 projections were collected with 5 s exposure each, at a 0.15\textdegree\ spacing, resulting in an isotropic voxel (pixel) size of 6.26 \textmu m. The Zeiss Scout-and-Scan software was used to reconstruct projects into 3D datasets, and binary images were generated by applying thresholding (Otsu, global) using Fiji \cite{Schindelin2012}. The software package StrutSurf \cite{Oosterbeek2022} was used to analyse micro-CT data and measure strut properties, using a 1.352 mm region size for strut selection. Elliptical cylinder geometry was fitted to calculate strut diameter (area equivalent circle diameter $D_{eq}$),  surface height plots, and surface roughness parameters (average roughness $Sa$ and peak-to-valley height $Sz$). 20 struts were measured for each sample, distributed across the sample volume.

Optical microscopy was carried out using a Hirox RH-2000 digital microscope with MXB-5000REZ lens at 140, 700, and 4000$\times$ magnification (1.13, 0.23, and 0.04 \textmu m pixel resolution). 3D multi-focus images were obtained using the Hirox RH-2000 control software. Scanning electron microscopy (SEM) was performed using a TESCAN Mira (TESCAN, Czech Republic) in backscattered and secondary electron imaging modes, beam voltage and current of 10 keV and 3 nA.

\subsection{Quasi-static mechanical testing}
Quasi-static compression testing of lattice samples was completed in accordance with ISO 13314 \cite{ISO13314} using an Instron 5565 test frame with a 5 kN load cell. Samples were centred between two hardened and lubricated steel platens, and compression was carried out at 2 mm/min. Sample displacement was measured by averaging the signal from two LVDTs (RDP D6/05000A, RDP Electronics Ltd., UK). One preliminary test was carried out for each unique sample to estimate the ultimate strength, by loading up to 50\% strain. Five samples were then tested by loading up to 50\% strain to measure the mechanical properties. A hysteresis loop was used to measure the elastic modulus during these tests, reversing from 70\% of the estimated ultimate strength down to 20\%, before continuing loading up to 50\% strain. A linear slope was fitted to the hysteresis loop to calculate the elastic modulus ($E$), and the yield strength ($\sigma_{y}$) was determined at a 1\% offset from this, as shown in Fig. \ref{fig:QS_mech}.

\subsection{Fatigue testing}
Compression-compression fatigue testing was carried out in accordance with ISO 1099 \cite{ISO1099}. A servo-hydraulis Instron 8872 load frame was used, fitted with either a 1, 5, or 25 kN dynamic load cell depending on the sample mechanical properties. Samples were held between two hardened and lubricated steel platens. Load control was used to determine crosshead displacement, with a programmed sinusoidal load waveform at a frequency of 15 Hz. A stress ratio ($R$) of 0.1 (min:max compressive stress ratio) was used.  Tests were stopped when either 2 mm displacement was reached ($\sim$10\% strain), or when the total number of cycles reached $10^{6}$. The fatigue life for a given fatigue stress level was taken as the number of cycles at 5\% nominal strain. Two specimens per set of samples were tested at each of 5 stress levels (80, 65, 50, 35, and 25\% of $\sigma_{y}$) to generate an \textit{S-N} curve. The modified staircase method was used as per ISO 12107 \cite{ISO12107} to determine the fatigue strength at $10^{6}$ cycles using an estimated initial stress and a stress step of 1.25\% of $\sigma_{y}$, continuing until 6 valid tests were completed.

\section{Results}

\FloatBarrier

\subsection{Morphological changes}
\begin{figure}[tbp]
	\centering
	\includegraphics[width=90mm]{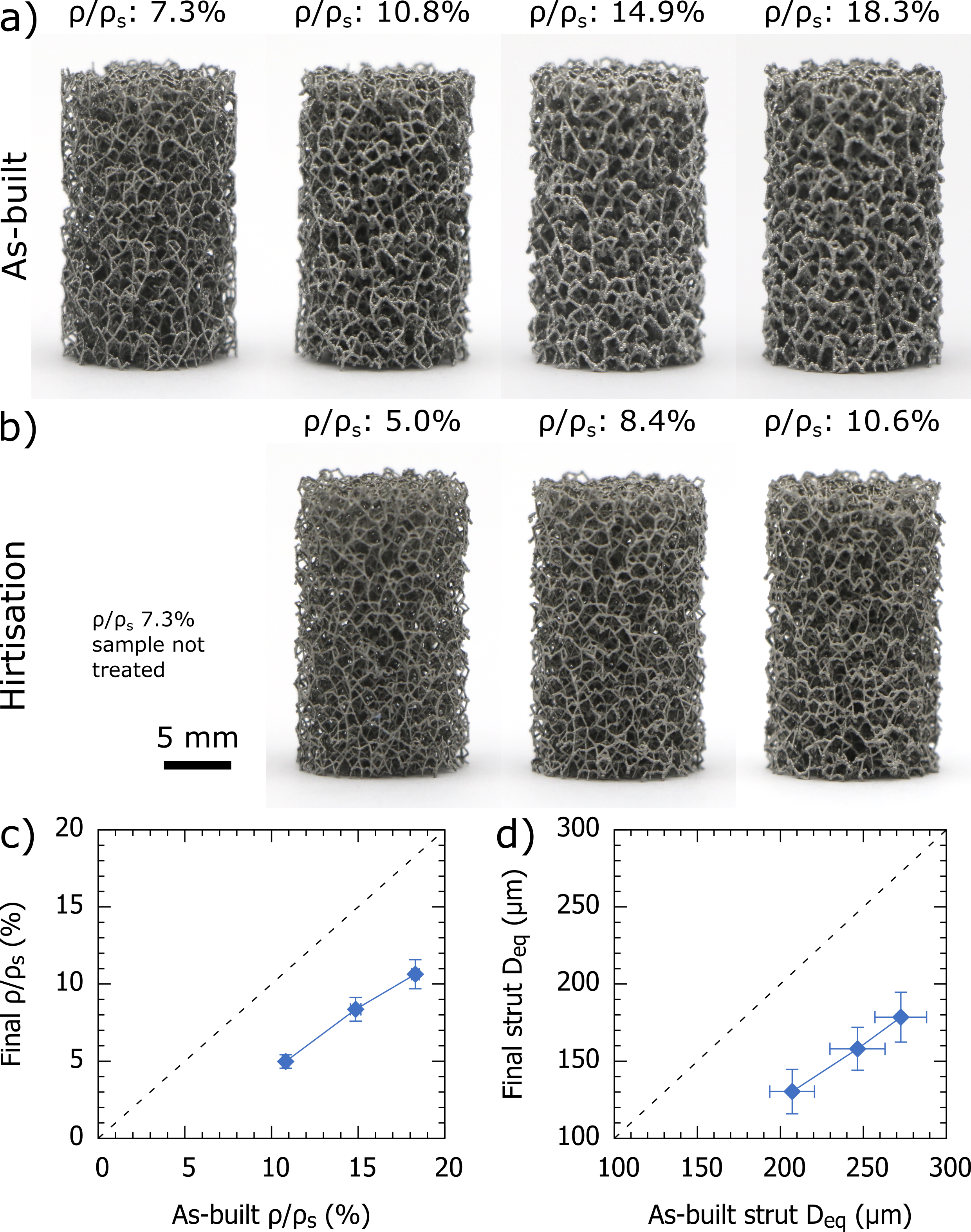}
	\caption{Photographs of lattices in the as-built condition (a) and after hirtisation (b). Graphs (c) and (d) show correlations of relative density $\rho /\rho_{s}$ and strut diameter $D_{eq}$ respectively, in the as-built condition and after hirtisation, to enable correction for material loss in the design stage. Dotted lines indicate no change i.e. final = as-built.}
	\label{fig:Bulk_changes_FIGURE}
\end{figure}

At the scale of a whole lattice sample (\diameter13 $\times$ 22 mm) morphological changes were observed as a result of the hirtisation treatment. Most importantly, considerable material loss was found, with reduction in the relative density $\rho /\rho_{s}$ by 42-54\% as shown in Table \ref{tab:MorphQSMech}. This is a result of loss of strut thickness, as measured by micro-CT, demonstrating a reduction in strut diameter $D_{eq}$ by 34-37\%. These changes are summarised in Figure \ref{fig:Bulk_changes_FIGURE}, showing photographs of the lattices before and after hirtisation, along with the measured properties. Correlation plots in Figure \ref{fig:Bulk_changes_FIGURE}c-d of properties like relative density and strut diameter before and after hirtisation can be utilised in the lattice design stage to correct for this mass loss and fabricate lattices with desired dimensions after hirtisation.

\begin{table}[tb]
	\centering
	\caption{\label{tab:MorphQSMech}Summary of the lattice morphological properties and quasi-static mechanical properties in the as-built condition and after hirtisation. Values given as mean$\pm$standard deviation.}
	\begin{tabularx}{120mm}{X r r r r}
		\toprule
		& \multicolumn{4}{c}{\textbf{As-built}} \\
		\textbf{$\rho /\rho_{s}$ (\%)} & \textbf{7.3$\pm$0.1} & \textbf{10.8$\pm$0.2} & \textbf{14.9$\pm$0.3} & \textbf{18.3$\pm$0.2} \\
		\midrule
		$D_{eq}$ (\textmu m) & 186$\pm$9 & 207$\pm$13 & 246$\pm$17 & 273$\pm$16 \\
		$Sa$ (\textmu m) & 12.4$\pm$1.5 & 11.2$\pm$1.3 & 11.8$\pm$1.3 & 12.4$\pm$1.4 \\
		$Sz$ (\textmu m) & 105$\pm$17 & 98$\pm$14 & 112$\pm$23 & 109$\pm$24 \\
		$\sigma_{y}$ (MPa) & 2.4$\pm$0.1 & 7.9$\pm$0.3 & 17.4$\pm$0.5 & 25.2$\pm$0.3 \\
		$\sigma_{u}$ (MPa) & 2.6$\pm$0.1 & 8.3$\pm$0.3 & 19.7$\pm$1.1 & 28.7$\pm$0.5 \\
		$E$ (MPa) & 159$\pm$3 & 538$\pm$20 & 1209$\pm$54 & 1839$\pm$14 \\
		\midrule	
		& \multicolumn{4}{c}{\textbf{Hirtisation}} \\
		\textbf{$\rho /\rho_{s}$ (\%)} & & \textbf{5.0$\pm$0.4} & \textbf{8.4$\pm$0.8} & \textbf{10.6$\pm$0.9} \\
		\midrule	
		$D_{eq}$ (\textmu m) & & 130$\pm$14 & 158$\pm$14 & 179$\pm$16 \\
		$Sa$ (\textmu m) & & 5.8$\pm$0.7 & 5.4$\pm$1.0 & 5.8$\pm$0.9 \\
		$Sz$ (\textmu m) & & 50$\pm$7 & 50$\pm$14 & 61$\pm$15 \\
		$\sigma_{y}$ (MPa) & & 2.9$\pm$0.4 & 7.3$\pm$1.9 & 10.8$\pm$2.8 \\
		$\sigma_{u}$ (MPa) & & 3.1$\pm$0.5 & 7.9$\pm$2.2 & 11.5$\pm$3.2 \\
		$E$ (MPa) & & 155$\pm$30 & 510$\pm$41 & 802$\pm$140 \\
		\bottomrule
	\end{tabularx}
\end{table}

\subsection{Surface changes - qualitative}\label{subsec:SurfQual}
\begin{figure}[tbp]
	\centering
	\includegraphics[width=90mm]{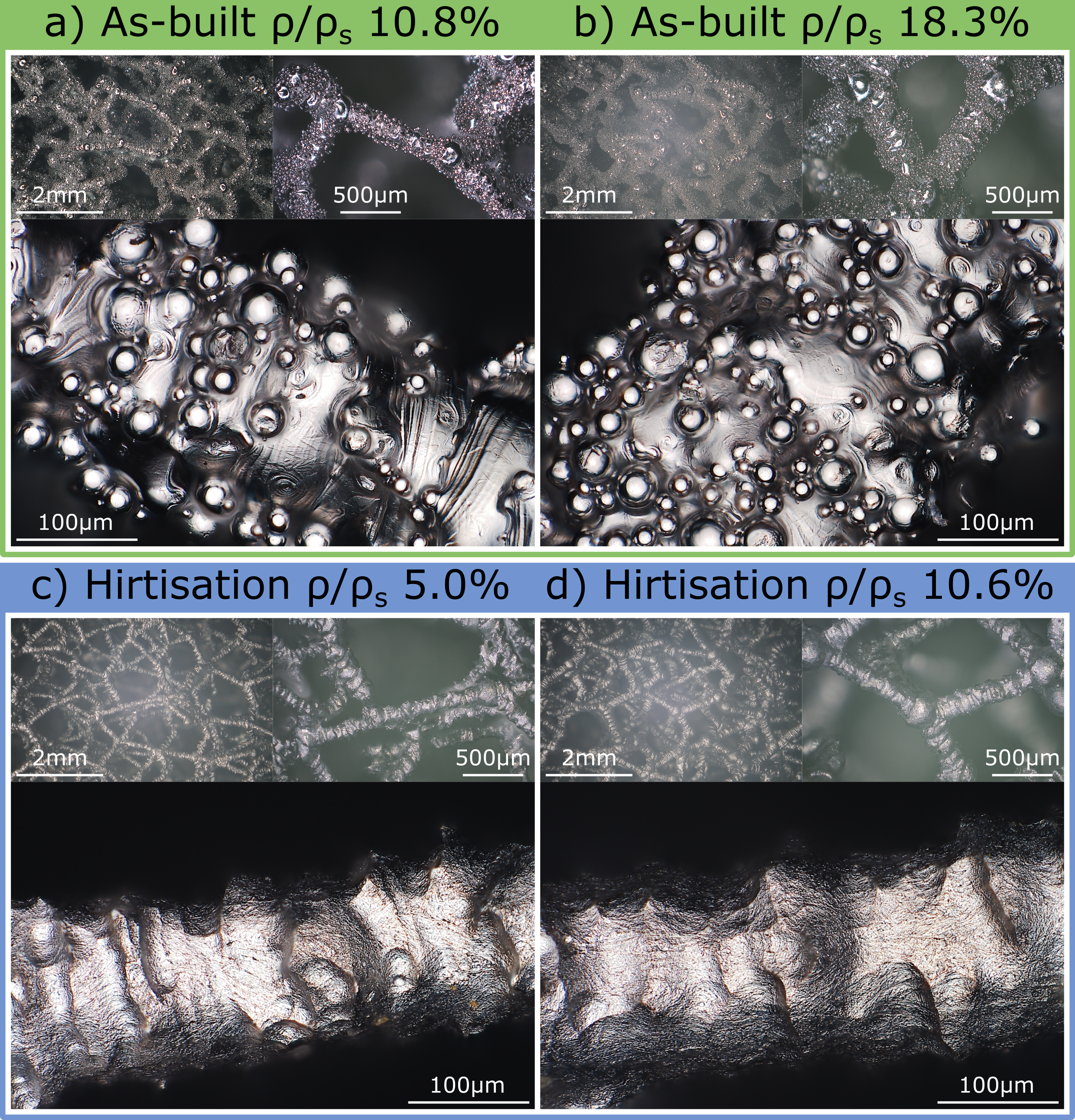}
	\caption{Multi-focus optical microscope images of lattices and struts in the as-built condition: (a) $\rho /\rho_{s} =$ 10.8\%, (b) $\rho /\rho_{s} =$ 18.3\%, and after hirtisation: (c) $\rho /\rho_{s} =$ 5.0\%, (d) $\rho /\rho_{s} =$ 10.6\%. Build direction is horizontal, left to right.}
	\label{fig:Microscope_images}
\end{figure}

\begin{figure}[tbp]
	\centering
	\includegraphics[width=140mm]{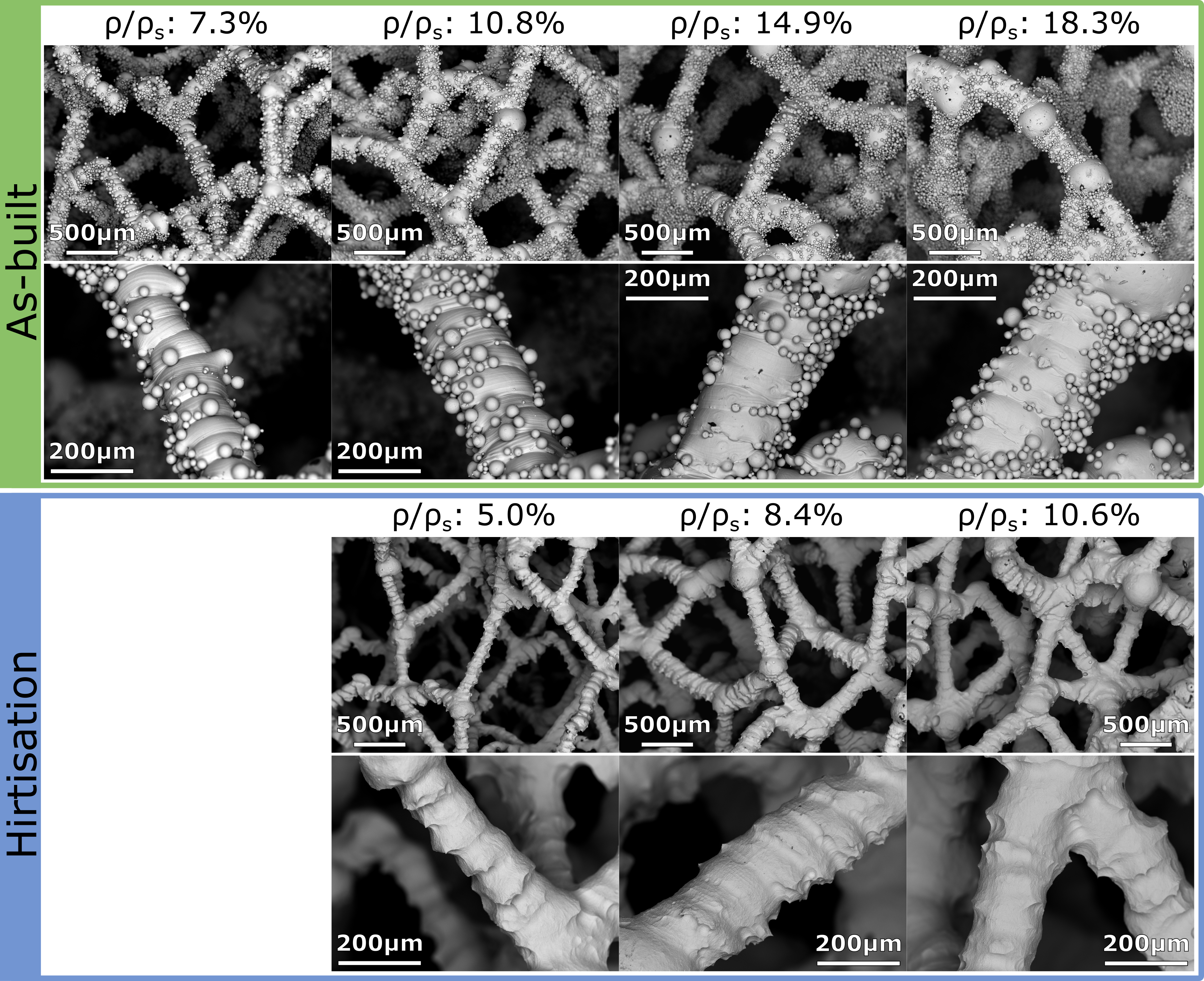}
	\caption{SEM images (in backscattered electron mode) of lattices and struts in the as-built condition and after hirtisation. Build direction is vertical, bottom to top.}
	\label{fig:SEM_images}
\end{figure}

A qualitative view of the changes in surface roughness as a result of the hirtisation process can be seen in Figures \ref{fig:Microscope_images} and \ref{fig:SEM_images}, showing optical microscope and SEM images respectively. In the as-built condition, lattice struts are characterised by many semi-fused particles attached to the surface, which have survived the ultrasonic  cleaning process. These are found all across the strut surface, but particularly on the underside of angled struts and underneath nodes due to the overhanging geometry. In addition to the semi-fused particles, surface roughness arising from weld lines between layers is clearly visible for all as-built struts.

The qualitative changes in the strut surface as a result of hirtisation are quite clear from microscope images in Figures \ref{fig:Microscope_images} and \ref{fig:SEM_images}. The hirtisation treatment has removed all the semi-fused particles from the strut surface, with none observed for any of the post-hirtisation samples. This is true for both the struts and nodes within the structure - we may expect semi-fused particles at nodes to be more difficult to remove, but this was not observed to be the case. Some surface roughness remains, with thickness variations apparent along the length of the strut with spacing on the order of $\sim$100 \textmu m (2 build layers) apart, consistent with the weld lines observed in the as-built struts. The geometry of the lattice structure also seems to have been well preserved, with the stochastic network still intact and the node/strut size ratio appearing reasonably unchanged.

\begin{figure}[tbp]
	\centering
	\includegraphics[width=90mm]{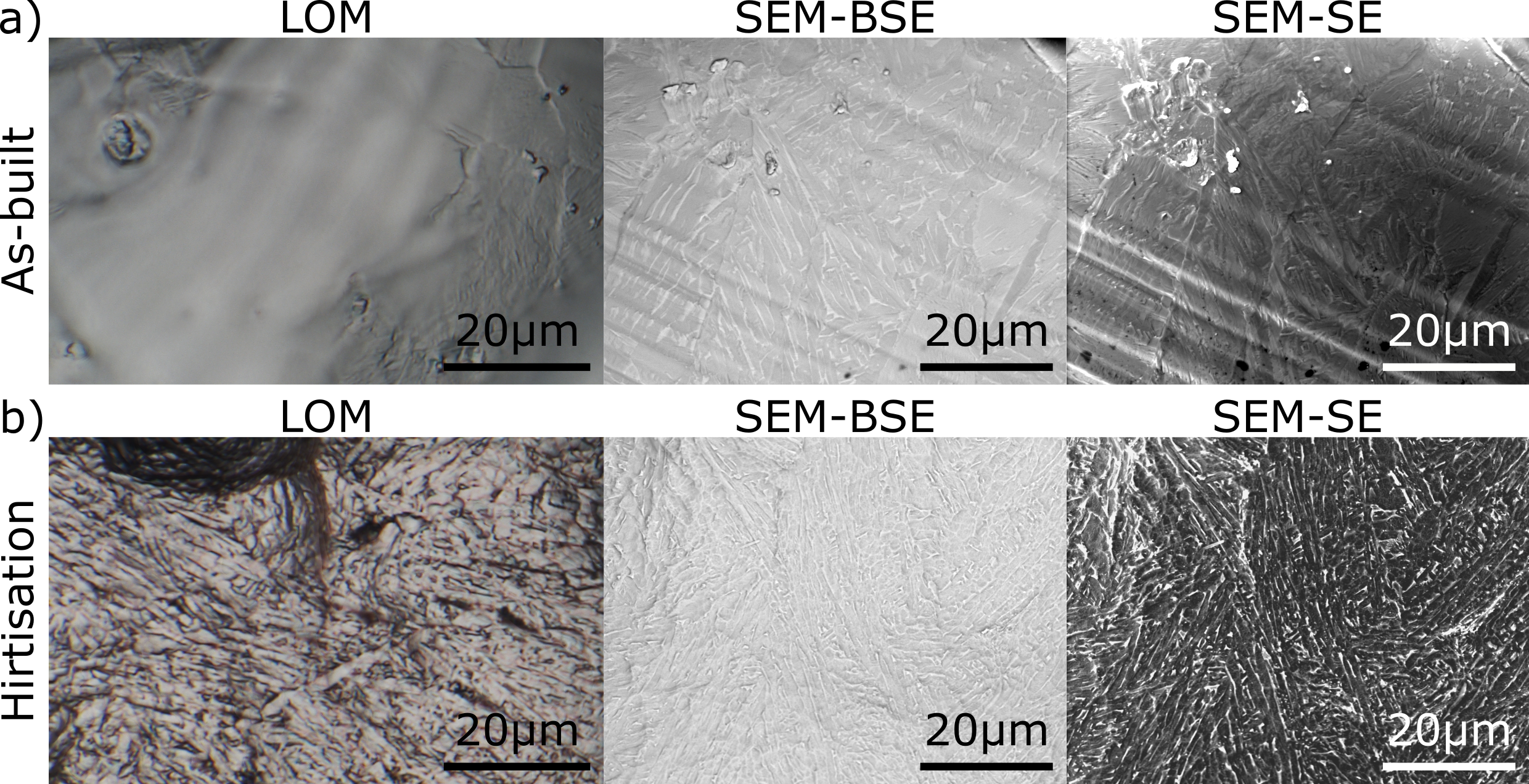}
	\caption{Images of the material microstructure as visible on the strut surface. Showing the as-built condition (a) and after hirtisation (b) by multi-focus light optical microscopy (LOM), SEM in backscatter electron mode (SEM-BSE) and SEM in secondary electron mode (SEM-SE).}
	\label{fig:Surface_microstructure}
\end{figure}

The aforementioned changes in surface roughness as a result of semi-fused particles and weld lines all occur on a scale of 20-100 \textmu m. However from high magnification optical and electron microscopy images (Figure \ref{fig:Surface_microstructure}) it can be seen that there are further changes in surface roughness on a scale of $\sim$1 \textmu m. These appear to be a result of the alloy microstructure, which is known to form a fine lamellar microstructure after sub-transus heat treatment as performed here \cite{Ghouse2021}. This microstructure is faintly visible on the surface of the as-built lattice struts, where lamellar features can be observed. After hirtisation, a process that includes electrochemical etching, this microstructure is more pronounced and visible on the surface, potentially due to slightly accelerated etching along grain boundaries.

\subsection{Surface changes - quantitative}\label{subsec:SurfQuant}
\begin{figure}[tbp]
	\centering
	\includegraphics[width=140mm]{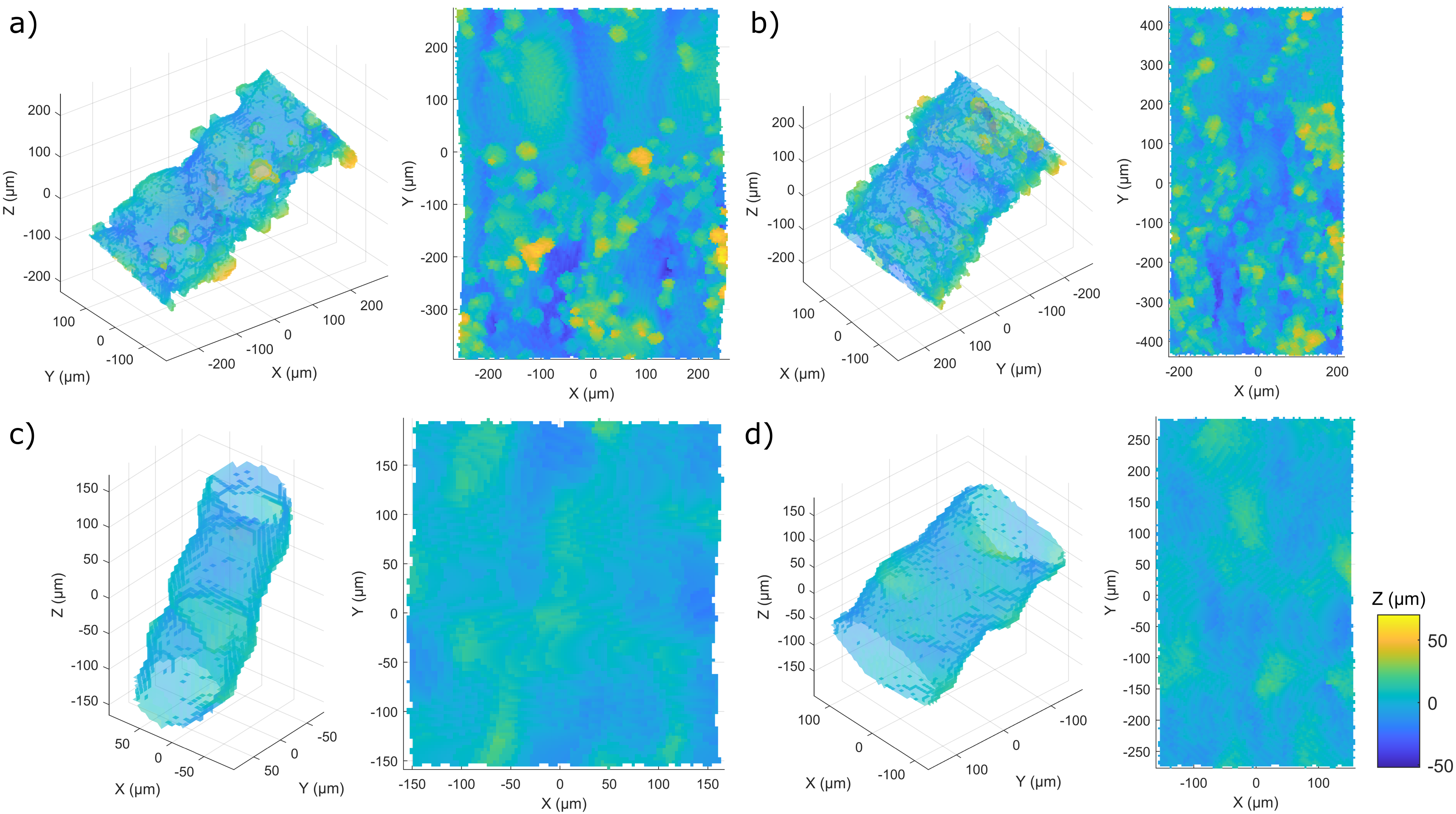}
	\caption{Micro-CT data analysed by StrutSurf, showing 3D strut surface plots and extracted surfaces for roughness calculation. All datasets have been plotted on the same colour scale. Showing strut surfaces in the as-built condition (a) $\rho /\rho_{s} =$ 10.8\%, (b) $\rho /\rho_{s} =$ 18.3\%, and after hirtisation: (c) $\rho /\rho_{s} =$ 5.0\%, (d) $\rho /\rho_{s} =$ 10.6\%.}
	\label{fig:MicroCT_images}
\end{figure}

Imaging results in section \ref{subsec:SurfQual} provide a qualitative view of the strut surface roughness, demonstrating the reduction in surface roughness achieved by hirtisation. However to comprehensively characterise the lattice roughness and the effectiveness of the treatment method, quantitative analysis is desirable. Figure \ref{fig:MicroCT_images} shows 3D images of individual struts obtained by micro-CT, along with their extracted surface plots, for lattice struts in the as-built condition and after hirtisation. Similar results are obtained to what can be seen in microscopy images (Figures \ref{fig:Microscope_images} and \ref{fig:SEM_images}). The as-built struts show high surface roughness due to the semi-fused particles that can be seen on the surface, as well as more gradual surface height changes from weld lines between different layers. After hirtisation all the semi-fused particles are removed from the strut surface, corroborating earlier microscopy results. As seen from microscopy, the remaining surface roughness consists of strut thickness variations along the strut length, with spacing on the order of $\sim$100 \textmu m, consistent with weld lines observed in the as-built condition. This indicates that although semi-fused particles are effectively removed from the lattice by hirtisation, completely smoothing the struts and removing weld lines between layers is more challenging and not achievable in this context. The fine surface texture on a $\sim$1 \textmu m scale that was observed by microscopy (Figure \ref{fig:Surface_microstructure}) could not be detected using micro-CT - this is a result of the pixel size used. In micro-CT the pixel size is directly linked to the field of view size, therefore to scan the whole 13 mm diameter of the sample (crucial for roughness distribution measurements, see below) the minimum pixel size achievable was 6.26 \textmu m. However, with average strut surface variations (peak-to-valley height $Sz$) on the order of 50-112 \textmu m (Table \ref{tab:MorphQSMech}), ignoring this fine surface texture is unlikely to significantly alter the results.

From the extracted surface plots, after fitting struts with elliptic cylinders, areal roughness parameters ($Sa$ and $Sz$) can be calculated for each strut. After measuring 20 struts for each lattice type (different relative densities, in the as-built condition and after hirtisation) the average roughness $Sa$ and peak-to-valley height $Sz$ are shown in Table \ref{tab:MorphQSMech}. These values provide quantitative information to demonstrate the reduction in surface roughness observed qualitatively by imaging methods. For these lattices the hirtisation process is seen to reduce both measures of roughness by approximately 50\%, on average reducing $Sa$ from $\sim$12 \textmu m to $\sim$6 \textmu m and $Sz$ from $\sim$110 \textmu m to $\sim$54 \textmu m.

\begin{figure}[tbp]
	\centering
	\includegraphics[width=90mm]{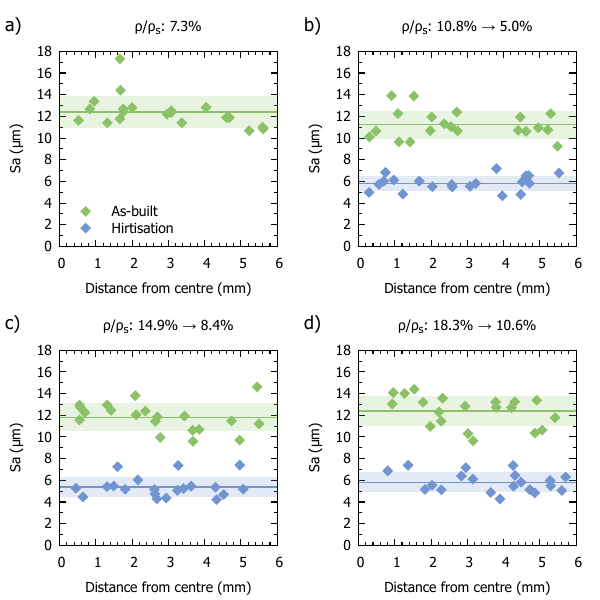}
	\caption{Variations in strut surface roughness across the radial distance from the lattice centre for struts within a \diameter13 mm lattice, in the as-built condition and after hirtisation. Lines and shaded regions show the average and standard deviation of the strut surface roughness, $n = 20$. $\rho /\rho_{s}$ values are given for lattices in the as-built condition and after hirtisation.}
	\label{fig:Roughness}
\end{figure}

In addition to average surface roughnesses for an entire lattice, measuring the surface roughness of individual struts can provide valuable information about roughness variation throughout a lattice, important for determining the effectiveness of surface treatment techniques. The average surface roughness $Sa$ for each strut measured is shown in Figure \ref{fig:Roughness} as a function of its radial distance from the centre of the lattice. As the hirtisation process must penetrate the lattice from the outside in, this gives an indication of the effectiveness and evenness of the surface treatment throughout the lattice. The roughness of the lattice struts after hirtisation does not appear to show any dependence on the position within the lattice. This was confirmed by linear regression, with $p$ = .385, .809, and .203 for $\rho /\rho_{s}$ = 5.0\%, 8.4\%, and 10.6\% respectively, indicating no evidence against the null hypothesis that $Sa$ and the radial distance from the lattice centre are not correlated. Simlar results were also seen for the strut diameter $D_{eq}$. This indicates that the hirtisation process does not unevenly etch lattice struts, but rather provides consistent surface treatment throughout these 13 mm \diameter\ lattices.

\subsection{Quasi-static mechanical properties}
\begin{figure}[tbp]
	\centering
	\includegraphics[width=90mm]{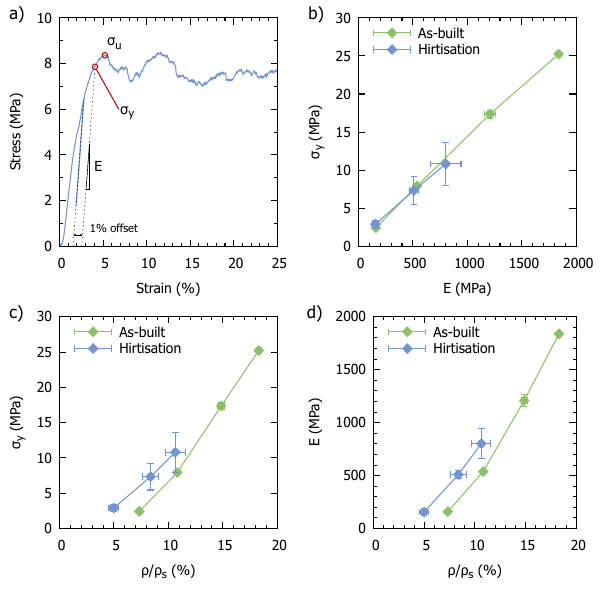}
	\caption{Quasi-static mechanical properties of lattices in the as-built condition and after hirtisation, showing (a) an example stress-strain curve for a lattice (as-built, $\rho /\rho_{s} =$ 10.8\%) to illustrate how $E$, $\sigma_{y}$, and $\sigma_{u}$ were determined, along with Ashby plots showing (b) $\sigma_{y}$ vs. $E$, (c) $\sigma_{y}$ vs. $\rho /\rho_{s}$, and (d) $E$ vs. $\rho /\rho_{s}$ for the as-built condition and after hirtisation.}
	\label{fig:QS_mech}
\end{figure}

Results of quasi-static mechanical testing are shown in Figure \ref{fig:QS_mech}. The yield strength $\sigma_{y}$ and elastic modulus $E$ are seen to increase with relative density $\rho /\rho_{s}$ as expected (Figure \ref{fig:QS_mech}c-d), with the as-built lattices displaying comparable mechanical properties to those observed by Hossain et al. for lattices with similar structure in the same material (Ti-6Al-4V) \cite{Hossain2021}. After hirtisation, the dependence of mechanical properties ($\sigma_{y}$ and $E$) on relative density $\rho /\rho_{s}$ appears similar, with the curve shifted towards lower relative density. This indicates that lattices treated using hirtisation can achieve similar mechanical strength at lower density compared with as-built lattices. This can be easily understood with reference to the effect of the hirtisation process, which is observed to remove material from the outside of the strut and reduce surface variations. In particular features such as semi-fused particles are eliminated, which would have negligible contribution to the load carrying ability and mechanical properties, but would contribute to the overall lattice density.

Figure \ref{fig:QS_mech}b shows an Ashby plot of the yield strength and modulus for lattices in the as-built condition and after hirtisation. The $\sigma_{y}/E$ ratio appears the same in both cases, suggesting that changes in quasi-static mechanical properties are purely due to material loss and not a result of any changes to the material itself.

\subsection{Fatigue properties}
\begin{figure}[tbp]
	\centering
	\includegraphics[width=140mm]{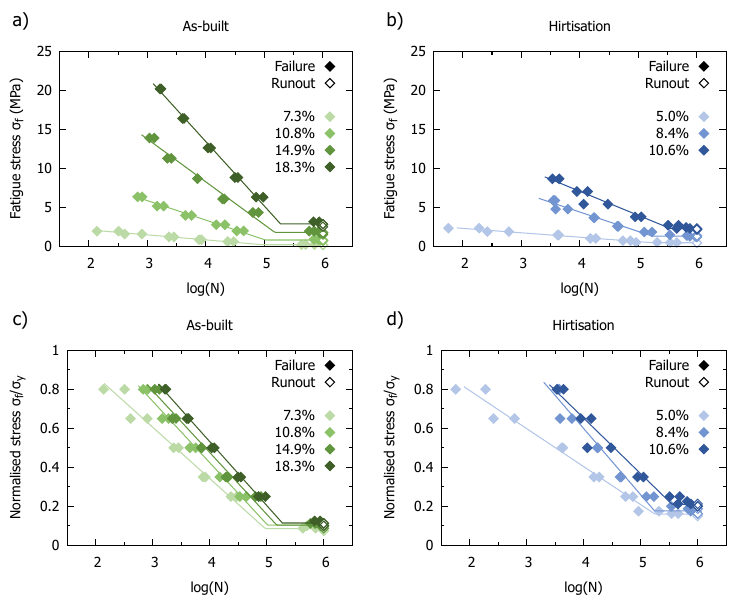}
	\caption{\textit{S-N} curves for compressive fatigue of lattices with different relative densities in the as-built condition (a, c) and after hirtisation (b, d), showing peak fatigue stress (a, b) and peak fatigue stress normalised to the quasi-static yield strength (c, d). Data points reaching $10^{6}$ cycles without failure (runout) during the modified staircase method are denoted with empty symbols. Lines shown are the fitted model according to Eq. \ref{eq:fatigue} with fitted parameters in Table \ref{tab:Fatigue}.}
	\label{fig:Fatigue_SN}
\end{figure}

Plots of peak test stress $\sigma_{f}$ against cycles to failure (\textit{S-N} curves) are shown in Figure \ref{fig:Fatigue_SN}, for lattices with different relative densities in the as-built condition and after hirtisation. Data are shown fitted to the model described in ISO 12107 \cite{ISO12107}, for the \textit{S-N} curve in both the finite and infinite fatigue life regions:

\begin{equation}\label{eq:fatigue}
	\begin{cases}
		\log N = \hat{b} - \hat{a}\sigma_{f} & \text{when } \sigma_{f} > \sigma_{f,10^{6}}\\
		\sigma_{f} = \sigma_{f,10^{6}} & \text{otherwise} 
	\end{cases}
\end{equation}

This model fits the observed data reasonably well in both the finite and infinite fatigue life regions, with fitted parameters given in Table \ref{tab:Fatigue}. Unsurprisingly reductions in the relative density of the lattice are seen to reduce the fatigue strength, however even when normalised to the static yield strength $\sigma_{y}$ these differences persist, with lower density lattices failing at fewer cycles for the same normalised fatigue stress. This indicates that lower relative density lattices are more sensitive to fatigue failure, which could be explained by their increased surface area to volume ratio (reduced strut thickness), magnifying the effect of surface flaws on the fatigue process. 

\begin{table}[tbp]
	\centering
	\caption{\label{tab:Fatigue}Summary of the high cycle ($10^{6}$ cycles) fatigue properties and fitted parameters for the mathematical model (Eq. \ref{eq:fatigue}) for lattices in the as-built condition and after hirtisation. Values given as mean$\pm$standard deviation.}
	\begin{tabularx}{140mm}{X r r r r}
		\toprule
		& \multicolumn{4}{c}{\textbf{As-built}} \\
		\textbf{$\rho /\rho_{s}$ (\%)} & \textbf{7.3$\pm$0.1} & \textbf{10.8$\pm$0.2} & \textbf{14.9$\pm$0.3} & \textbf{18.3$\pm$0.2} \\
		\midrule
		$\hat{a}$ (MPa\textsuperscript{-1}) & 1.5612 & 0.4002 & 0.1820 & 0.1210 \\
		$\hat{b}$ & 5.3019 & 5.3463 & 5.5041 & 5.6244 \\
		$\sigma_{f,10^{6}}$ (MPa) & 0.21$\pm$0.09 & 0.82$\pm$0.21 & 1.79$\pm$0.64 & 2.89$\pm$0.48 \\
		$\sigma_{f,10^{6}}/\sigma_{y}$ & 0.085$\pm$0.037 & 0.103$\pm$0.027 & 0.103$\pm$0.037 & 0.115$\pm$0.019 \\
		$\sigma_{f,10^{6}}/E$ ($\times10^{3}$) & 1.31$\pm$0.57 & 1.52$\pm$0.40 & 1.48$\pm$0.53 & 1.57$\pm$0.26 \\
		\midrule	
		& \multicolumn{4}{c}{\textbf{Hirtisation}} \\
		\textbf{$\rho /\rho_{s}$ (\%)} & & \textbf{5.0$\pm$0.4} & \textbf{8.4$\pm$0.8} & \textbf{10.6$\pm$0.9} \\
		\midrule	
		$\hat{a}$  (MPa\textsuperscript{-1}) & & 1.7632 & 0.4008 & 0.3240 \\
		$\hat{b}$ & & 6.0708 & 5.7643 & 6.2872 \\
		$\sigma_{f,10^{6}}$ (MPa) & & 0.47$\pm$0.11 & 1.30$\pm$0.38 & 2.28$\pm$0.66 \\
		$\sigma_{f,10^{6}}/\sigma_{y}$ & & 0.160$\pm$0.044 & 0.177$\pm$0.069 & 0.210$\pm$0.082 \\
		$\sigma_{f,10^{6}}/E$ ($\times10^{3}$) & & 3.01$\pm$0.91 & 2.55$\pm$0.77 & 2.84$\pm$0.96 \\
		\bottomrule
	\end{tabularx}
\end{table}

\begin{figure}[h]
	\centering
	\includegraphics[width=90mm]{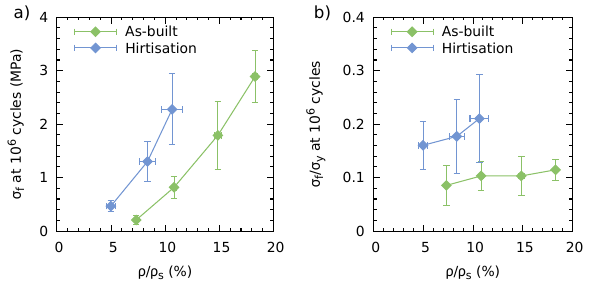}
	\caption{Ashby plots showing (a) the high cycle ($10^{6}$ cycles) fatigue strength vs. $\rho /\rho_{s}$, and (b) high cycle fatigue strength normalised to $\sigma_{y}$ vs. $\rho /\rho_{s}$, for both the as-built condition and after hirtisation.}
	\label{fig:Fatigue_trend}
\end{figure}

It can also be seen from these graphs that the normalised high cycle fatigue strength $\sigma_{f,10^{6}}/\sigma_{y}$ appears to have increased after hirtisation, though this is more obvious in Figure \ref{fig:Fatigue_trend}b. Here the normalised high cycle fatigue strength can be seen to have increased from around 0.1 to 0.16-0.21, an average increase of about 80\%. Figure \ref{fig:Fatigue_trend}a shows the variation in fatigue strength with relative density. In a similar way to the static yield strength (Figure \ref{fig:QS_mech}c) this curve is shifted to the left due to loss of material (such as semi-fused particles) that does not contribute to the load bearing capability, however the reduction in the fatigue strengths achieved is much lower than the reduction in yield strength. This demonstrates that reduction in $\sigma_{f}$ due to material loss is being offset by improvements in fatigue resistance caused by reduced surface roughness.

\begin{figure}[h]
	\centering
	\includegraphics[width=90mm]{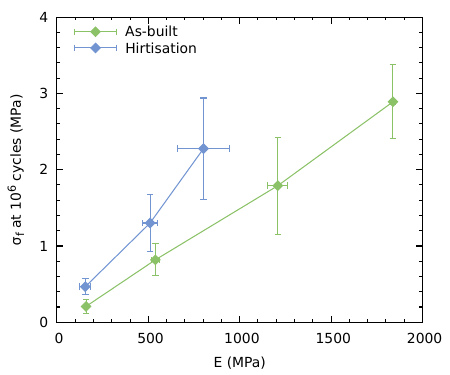}
	\caption{Ashby plot showing the high cycle ($10^{6}$ cycles) fatigue strength vs. modulus $E$ for the as-built condition and after hirtisation - a crucial comparison for maximising the fatigue resistance of stiffness-matched medical implant materials.}
	\label{fig:Fatigue_E}
\end{figure}

The correlation between high cycle fatigue strength and elastic modulus is shown in Figure \ref{fig:Fatigue_E}. Unlike for quasi-static conditions where the relationship of yield strength with $E$ does not appear to change after hirtisation, under dynamic conditions the dependence of fatigue strength on $E$ has distinctly changed after hirtisation. The increased $\sigma_{f,10^{6}}/E$ ratio therefore cannot be simply explained by removal of material that does not contribute to the load-bearing capability, indicating that the decreased roughness has reduced the crack initiation sites, thereby improving fatigue resistance. The $\sigma_{f,10^{6}}/E$ ratio is a key design parameter for orthopaedic lattice materials, and maximising this ratio will allow implant stiffness to be matched to that of natural bone, while achieving maximum fatigue strength for optimal implant lifetime. From Figure \ref{fig:Fatigue_E} it is clear that a lattice material designed to achieve a certain $E$ value based on the hirtisation process will have a markedly higher fatigue strength than one based on the as-built condition. This is quantified by the $\sigma_{f,10^{6}}/E$ ratio, listed in Table \ref{tab:Fatigue}, showing that the hirtisation process increased this ratio from an average of $1.47\times 10^{-3}$ to $2.8\times 10^{-3}$, an increase of over 90\%.

\section{Discussion}

\subsection{Effect of hirtisation on surface roughness}

The results of this work demonstrate that the roughness of Ti-6Al-4V lattices with $\rho /\rho_{s}$ up to 18.3\% can be reduced by approximately 50\% by hirtisation. This is slightly less than the approximately 75\% reduction observed by Sandell et al for hirtisation of bulk solids \cite{Sandell2022} due to the challenges of effectively treating internal surfaces. Previous works using simple chemical etching have demonstrated that etching methods can have some difficulties penetrating into a porous structure, with Ahmadi et al. speculating that this may be a result of trapping of gas bubbles formed during etching \cite{Ahmadi2019, Wysocki2019}. A comprehensive comparison to literature is difficult, as measurement of surface roughness changes within a porous lattice can be challenging and is therefore not frequently carried out. However, the reduction in roughness seen here is greater than that observed by de Formanoir et al. \cite{DeFormanoir2016} by HF and HNO\textsubscript{3} etching (28\% $R_{a}$ reduction), and more comparable to the observations of Pyka et al. \cite{Pyka2012a} by combined chemical etching and electrochemical polishing (up to 50\% $P_{a}$ reduction observed).

Comprehensive characterisation of the effects of surface treatment methods is crucial to ensuring the safety and integrity of components processed using these methods. This should include not just roughness measurements, but assessment of the surface roughness throughout the porous structure, to reveal any changes in treatment efficacy deep within the lattice. Due to the technical challenges of measuring internal surface roughness however, this is rarely carried out in the literature. After etching de Formanoir et al. measured $R_{a}$ values of 25.9 \textmu m and 23.8 \textmu m for internal and external struts respectively (fabricated by electron beam melting), concluding that the treatment was homogeneous \cite{DeFormanoir2016}, while for electropolishing Kerckhofs et al. measured $P_{a}$ values of 8 \textmu m and 5 \textmu m for internal and external struts respectively, concluding that further optimisation of their electrochemical polishing settings was necessary \cite{Kerckhofs2012}. By utilising micro-CT and data analysis methods \cite{Oosterbeek2022} here we have assessed the resulting strut surface roughness as a function of the distance from the lattice centre (i.e. the most difficult portion for treatment to penetrate). We observed no substantial changes in the resulting surface roughness across the lattice, indicating that the hirtisation settings used were well optimised for providing even surface treatment. In addition to demonstrating the effectiveness of the hirtisation technique, we hope that this work will encourage more comprehensive evaluation of surface treatment methods in the future, making using of modern micro-CT and data analysis techniques.

A major motivation for these efforts to improve lattice surface roughness is the fatigue behaviour, where reduction in surface flaws and roughness is expected to improve the fatigue strength and lifetime of the lattice \cite{Karami2020, Ahmadi2019, Persenot2020}. This will be discussed in detail in section \ref{subsec:discuss-fatigue}, however the effect on the biological performance is also useful to consider. Semi-fused metal particles on the lattice strut surfaces, being poorly bonded to the main strut, may pose a risk of particle release. This release of particles containing metals like titanium and in vanadium may cause adverse effects \textit{in vivo}. Current literature on the effects of these particles is not yet comprehensive, however some studies indicate that metallic particles may increase bone resorption (osteolysis), a leading cause of implant failure \cite{Cunningham2003, Wang1996, Sansone2013}. There is also some evidence that metallic particles can trigger an adverse immune response, which may lead to pain and clinical failure \cite{Goodman2007}. In an oral context release of titanium particles has been linked to peri-implantitis, which involves soft tissue inflammation and bone loss and can lead to implant failure \cite{Kheder2021}. Removal of these particles by surface treatments such as hirtisation, as evidenced in Figures \ref{fig:Microscope_images} and \ref{fig:SEM_images}, is expected to drastically reduce the probability of particle release from an additively manufactured implant \textit{in vivo}, avoiding some of these potential hazards. 

Despite the observed improvements, this study does have some limitations with regards to characterisation of the effect of hirtisation on the surface roughness of lattice struts. One important limitation to consider is the pixel size used in micro-CT analysis of strut roughness - 6.26 \textmu m. While modern micro-CT systems such as that used here are capable of sub-micron resolution, this comes at the cost of reduced field-of-view (FOV). In this study chief importance was placed on obtaining larger FOV images in order to monitor any variation in treatment effectiveness throughout the sample, thus necessitating a larger pixel size. Nevertheless, we do not believe this will have had a substantial effect on the results obtained, as the main features contributing to the surface roughness (semi-fused particles, weld lines) can be clearly resolved at the pixel size used (see Figure \ref{fig:MicroCT_images}). This is further confirmed by comparison to similar lattice struts in the literature produced by laser powder bed fusion, which display comparable average roughness values before and after etching to those obtained here \cite{Pyka2012a, Pyka2013, Hossain2021a}. Additional surface features on the 1 \textmu m scale arising from the material microstructure were observed by microscopy, which were not captured by micro-CT, however the fine scale of these features means they are not expected to impact the average surface roughness. Another limitation is the range of relative density values and dimensions of the as-built lattices investigated in this study. We did not observe any reduction in the effectiveness of the hirtisation treatment deep within the lattice, however at higher relative densities or for larger lattices this effect may become more significant. We therefore cannot discount a reduction in surface treatment effectiveness for lattice relative densities greater than $\rho /\rho_{s} =$ 18.3\%, or for  a required treatment depth greater than 6.5 mm.

\subsection{Effect of hirtisation on fatigue performance}\label{subsec:discuss-fatigue}

The effect of hirtisation and the associated roughness reduction on the fatigue strength is quite clear. Due to material loss during the process it is most useful to consider normalised fatigue strength values ($\sigma_{f}/\sigma_{y}$), which increase by 80\% on average after hirtisation. These values and the increases observed are comparable if slightly lower than those reported by Ahmadi et al. and Karami et al., \cite{Ahmadi2019, Karami2020}. This difference can be attributed partly to the different lattice structure used, as well as the lower range of relative densities investigated here, which is observed to reduce the fatigue strength even after normalisation to the static yield strength (Figure \ref{fig:Fatigue_SN} and Table \ref{tab:Fatigue}). These works also utilise hot isostatic pressing (HIP) to provide further improvements to fatigue strength, however in this work we have only investigated the effects of surface treatment. Heat treatments such as super-transus vacuum heat treatment or HIP are known to improve the fatigue strength of AM Ti64 lattices through changes in microstructure and micro-morphology \cite{Ghouse2021, Ahmadi2019, Karami2020}, and in the future these could be combined with hirtisation to provide optimal resistance to fatigue failure.

A crucial challenge for successful application of additively manufactured lattices to orthopaedic implants is achieving sufficiently high fatigue strength to ensure a safe useful life, while simultaneously matching the stiffness of the surrounding bone (approximately 500 - 1000 MPa for trabecular bone). These two requirements can often be contradictory, thus maximising the $\sigma_{f}/E$ ratio is a key material design goal. From Figure \ref{fig:Fatigue_E} and Table \ref{tab:Fatigue} it is seen that hirtisation increases this ratio by approximately 90\%. In the context of an orthopaedic implant this means that an implant designed to match the stiffness of bone at a particular site will have a 90\% greater fatigue strength when designed based on the hirtisation process compared with the as-built lattice. Despite its importance as a design parameter the $\sigma_{f}/E$ ratio is not often reported in the literature, and we hope that this work will encourage use of this criteria as well as demonstrating an effective way to improve it.

One limitation of this work is the error size observed, particularly for the fatigue strength. Some of this can be attributed to increased variation in the relative density as a result of the hirtisation process, which can be seen in Table \ref{tab:MorphQSMech} and Figure \ref{fig:Bulk_changes_FIGURE}. The hirtisation procedure used provides consistent control of material removal, however it is still less consistent than the laser PBF process used for initial fabrication. This resulted in slight increases in the variation of relative density, which have carried through to the quasi-static mechanical properties. This suggests that there is still room for further optimisation of the specific hirtisation process used, to provide more consistent material removal. The other major cause of the error size observed for the fatigue strength is the measurement method used - the modified staircase method according to ISO 12107 \cite{ISO12107}. This method has been utilised in previous work \cite{Ghouse2021, Kechagias2022} and enables statistical estimation of the full \textit{S-N} curve (in both finite and infinite life regions) using fewer specimens than the full staircase method. The standard deviation of the fatigue strength is calculated from the finite life region of the \textit{S-N} curve, however this may overestimate the actual standard deviation of the high cycle fatigue strength $\sigma_{f,10^{6}}$. From the measurements in this work we can estimate that use of the full staircase method would reduce the observed standard deviation of the $\sigma_{f}/\sigma_{y}$ value by approximately 50\% on average. Due to time and material constraints the full staircase method could not be utilised for this exploratory study, however this may be beneficial for future tests with a reduced number of variables.

\section{Conclusions}

Hirtisation was able to effectively reduce the surface roughness of additively manufactured porous Ti-6Al-4V lattices, removing all visible semi-fused particles and reducing strut thickness variation. Through the use of 3D tomography (micro-CT), a roughness reduction of approximately 50\% was quantified, with \textit{Sa} reducing from around 12 to 6 \textmu m. The evenness of this treatment was also investigated, finding no evidence of reduced treatment effectiveness for lattice relative densities up to 18.3\% and treatment depths up to 6.5 mm. Higher relative densities and treatment depths were not included in this study. This roughness reduction is accompanied by a decrease in strut diameter by 34-37\%, which should be accounted for in the lattice design stage to achieve desired final dimensions.

These improvements in surface roughness led to improved fatigue performance, after normalising to quasi-static mechanical properties to account for material loss. The normalised high cycle (10\textsuperscript{6} cycles) fatigue strength ($\sigma_{f,10^{6}}/\sigma_{y}$) increased on average by 80\% after hirtisation, due to the surface roughness changes reducing the crack initiation sites. The $\sigma_{f}/E$ ratio also increased by approximately 90\% after hirtisation. This is an important metric for maximising the longevity of stiffness-matched medical implant lattices, and demonstrates the effectiveness of this method for orthopaedic implant devices.

\section*{CRediT authorship contribution statement}
\textbf{R.N. Oosterbeek}: Conceptualisation, Methodology, Formal analysis, Investigation, Visualisation, Writing. \textbf{G. Sirbu and S. Hansal}: Methodology. \textbf{K. Nai}: Methodology. \textbf{J.R.T. Jeffers}: Supervision, Funding acquisition.

\section*{Conflict of Interest}
G. Sirbu and S. Hansal are employees of Rena Technologies Austria GmbH and report a patent with Rena Technologies. The remaining authors have no conflicts of interest to declare.

\section*{Acknowledgements}
\label{}

The authors wish to gratefully acknowledge support from the Engineering and Physical Sciences Research Council (EP/R042721/1), and National Institute for Health Research (NIHR300013).



\bibliographystyle{ieeetr}

\bibliography{references.bib}






\end{document}